\pdfoutput=1
\documentclass[10pt,journal]{IEEEtran}
\usepackage{amsmath,amssymb,amsfonts}
\usepackage{mathtools}
\usepackage{dsfont}
\usepackage{soul}
\usepackage[dvipsnames]{xcolor}

\usepackage[export]{adjustbox}
\usepackage{caption}
\usepackage{algorithmic}
\usepackage{lettrine}
\usepackage{graphicx}
\usepackage{textcomp}
\usepackage{cancel}
\usepackage{hyperref}
\usepackage{subcaption}
\usepackage{float}
\usepackage[english]{babel}
\usepackage{amsthm}
\usepackage[ruled,vlined]{algorithm2e}
\usepackage[shortlabels]{enumitem}

\theoremstyle{definition}
\newtheorem{mydef}{Definition}
\newtheorem{myexam}{Example}
\DeclareGraphicsExtensions{.png,.pdf}
\usepackage{subcaption}
\usepackage{array}
\usepackage{multirow}
\usepackage{cite}
\usepackage{xcolor}
\usepackage{array,multirow,textcomp}
\theoremstyle{definition}

\usepackage{subcaption}
\begin{document}

\title{Intent-Aware DRL-Based NOMA Uplink Dynamic Scheduler for IIoT\\
\thanks{Salwa Mostafa is with the Centre for Wireless Communications, University of Oulu, FI-90014 Oulu, Finland and the Faculty of Electronic Engineering, Menofia University, Menofia, Egypt. Mehdi~Bennis is with the Centre for Wireless Communications, University of Oulu, FI-90014 Oulu, Finland (e-mail: salwa.mostafa, mehdi.bennis@oulu.fi). Mateus P. Mota is with the Institut National des Sciences Appliquées de Lyon (INSA Lyon) (e-mail:Mateus.pontes-mota@insa-lyon.fr). Alvaro Valcarce is with the Nokia Bell Labs, Nozay, France (e-mail:alvaro.valcarce\_rial@nokia-bell-labs.com). ({\em{Corresponding author: Salwa Mostafa.}}) \\
The simulation code for the paper is available on the following~\href{https://github.com/SalwaMostafa/Intent-Aware-DRL-Based-NOMA-Uplink-Dynamic-Scheduler-for-IIoT}{GitHub repository}.}}

\author{Salwa~Mostafa,~\IEEEmembership{Member,~IEEE}, Mateus P. Mota,~\IEEEmembership{Member,~IEEE}, Alvaro Valcarce,~\IEEEmembership{Senior Member,~IEEE}, and Mehdi~Bennis,~\IEEEmembership{Fellow,~IEEE}}

\maketitle

\begin{abstract}
We investigate the problem of supporting Industrial Internet of Things user equipment (IIoT UEs) with intent (i.e., requested quality of service (QoS)) and random traffic arrival. A deep reinforcement learning (DRL) based centralized dynamic scheduler for time-frequency resources is proposed to learn how to schedule the available communication resources among the IIoT UEs. The proposed scheduler leverages an RL framework to adapt to the dynamic changes in the wireless communication system and traffic arrivals. Moreover, a graph-based reduction scheme is proposed to reduce the state and action space of the RL framework to allow fast convergence and a better learning strategy. Simulation results demonstrate the effectiveness of the proposed intelligent scheduler in guaranteeing the expressed intent of IIoT UEs compared to several traditional scheduling schemes, such as round-robin, semi-static, and heuristic approaches. The proposed scheduler also outperforms the contention-free and contention-based schemes in maximizing the number of successfully computed tasks. 
\end{abstract}

\begin{IEEEkeywords}
IBN, IIoT, scheduling, mobile edge computing, reinforcement learning.
\end{IEEEkeywords}

\section{Introduction}

Wireless communication networks are expected to support User Equipments (UEs) with enhanced Mobile Broadband (eMBB), massive Machine-Type Communications (mMTC), and Ultra-Reliable Low-Latency Communications (URLLC) services~\cite{dahlman20205g}. The rapid growth in various technologies facilitate the implementation and support of Internet of Things applications and services, which motivates industrial operators to implement IoT in the industrial sectors known as the Industrial Internet of Things (IIoT)~\cite{garcia2021tutorial}. To allow IIoT UEs to express their various service demands in a simple way to the network, Intent-based networking (IBN) has been introduced, which provides a simple and efficient autonomic and autonomous way to configure and manage networks~\cite{leivadeas2022survey}. IBN relies on understanding what network users want and what are the network operator's capabilities to optimize the network operations with services. The UEs needs are expressed in an abstract and high-level language to the network. Therefore, {\em the intent is defined as a high-level and abstract description of the network services}~\cite{clemm2020intent}. In this work, we assume that IIoT UEs express their intent (i.e., requested QoS) through a graphical user interface (GUI) to the network~\cite{abbas2021network}.

However, IIoT has more demanding requirements to support machine devices with various services such as URLLC, where low latency and high reliability are essential to carry out critical real-time and interactive services such as remote surgery, intelligent transportation, autonomous control, and V2V communication~\cite{navarro2020survey,pokhrel2020towards}. Unfortunately, IIoT UEs have limited power and computation resources which limits their ability to support services with low latency and high reliability. Moreover, executing intensive computation tasks locally consumes high energy, which may exceed IIoT UEs's power capabilities. To tackle this problem, mobile edge computing (MEC), which provides high computation, control, storage, and communication resources on edge servers nearby IIoT UEs is incorporated into the IIoT architecture. This allows IIoT UEs to offload their computation-intensive tasks to edge servers for fast execution and power saving. However, offloading computation tasks to proximal edge serves adds an additional transmission delay~\cite{qiu2020edge,mao2017survey}. 

Therefore, efficient time-frequency communication resource scheduling schemes play a major role in reducing latency and guaranteeing reliability for IIoT UEs given the scarcity of radio resources. Thus, designing a scheduling scheme that takes into consideration different IIoT UE intents, gives an efficient utilization of network resources. Thus, in this work, we propose an intent-aware dynamic resource scheduler (DRS) based on an AI framework to learn an efficient scheduling strategy for IIoT UEs with various intents under random traffic arrival.

\subsection{Background}
The gNB scheduler decides when and which UEs are allowed to transmit/receive on the available resource blocks (RBs) and determines the transmission parameters. The gNB scheduler operates in two modes semi-static and dynamic in both the uplink and downlink directions. In the semi-static scheduling mode, the scheduled UEs and their transmission parameters are determined in advance. In contrast, in dynamic scheduling, the scheduler controls when and which UEs are assigned to the available RBs on a dynamic basis at each transmission time interval (i.e., time slot). In dynamic scheduling, scheduling decisions are made frequently, so variations in traffic demand and channel quality can be exploited to allocate available resources efficiently. 

The scheduler requires information about the channel quality, buffer status, and priority of data flow to make the scheduling decision. The gNB supports channel-dependent scheduling to obtain high multiuser diversity gain~\cite{chapman2014hspa}. The channel assignment is given over the control signaling in the same time slot in case of bandwidth adaptation. The gNB can easily obtain the channel conditions for UEs through CSI reports or channel reciprocity. However, the buffer status and traffic priorities are more challenging to obtain in uplink than downlink scheduling scenarios since UEs have limited power resources and are distributed (i.e., not residing on the same node (gNB) as in the downlink). 

\begin{figure} 
\centering
\includegraphics[width= 3 in, height = 3 in]{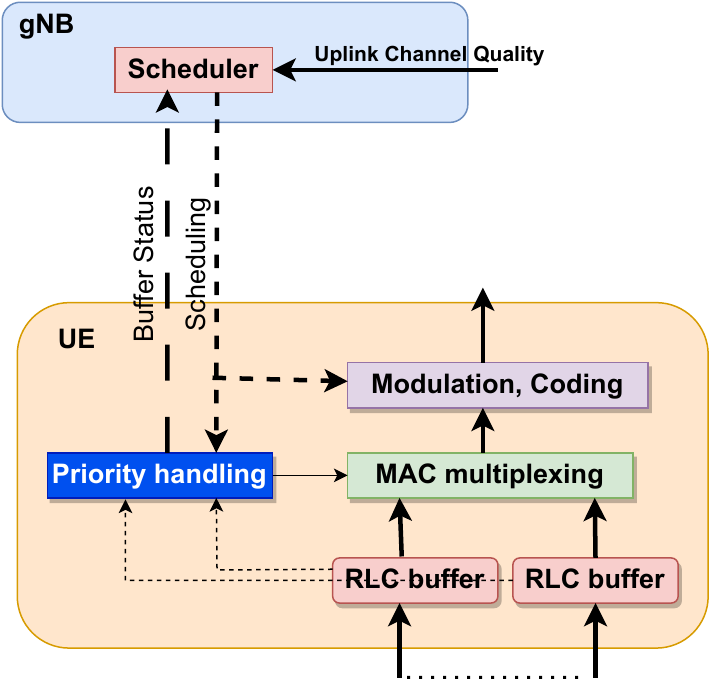}
\caption{Uplink dynamic scheduling.} 
\label{uplink}
\end{figure}

In dynamic scheduling, the scheduling strategy is not standardized but a set of supporting mechanisms are standardized to support the scheduling strategy implemented by a vendor. The gNB scheduler gives each UE a scheduling grant indicating the transmission resource block and transport format that shall be used, afterward, each UE follows the grant. The scheduler does the scheduling per device not per radio bearer so the logical channels are not explicitly scheduled but are provided to users according to rules to handle traffic priorities. Fig.~\ref{uplink} illustrates the uplink scheduling, where the scheduler controls the transport format and UEs are in charge of the logical-channel multiplexing. This provides a higher efficiency to resource usage than allowing each UE to autonomously control the transmission parameters in the system. 

The uplink scheduler is in charge of the transport format; thus, accurate information about each UE such as buffer status and power availability, must be provided to the scheduler. The buffer status report and power headroom report are sent through medium access control (MAC) control elements to gNB to decide the future resource allocation to each UE. If the UE is not assigned a scheduling grant and data traffic with high priority arrives, the UE has to send a scheduling request on the physical uplink shared channel (PUSCH) at the next possible instance to gNB. The UE informs the gNB that there is data that needs to be transmitted and about the type of data waiting for transmission. The scheduler supports dynamic TDD to dynamically determine the transmission direction as the half-duplex UEs cannot transmit and receive simultaneously, where there is a need to split the resources between the two directions~\cite{dahlman20205g}. 

\subsection{Related Work}

The orthogonal and non-orthogonal network slicing techniques are proposed in~\cite{guan2018service,zhang2017network,popovski20185g,yin2023connectivity} to support heterogeneous services such as mMTC and URLLC on the available radio resources. The studies in ~\cite{guan2018service,zhang2017network,popovski20185g} proposed network slicing approach treated the different services separately, which degrades the system performance due to the tradeoff between achieving the requirements of each service independently. The authors in~\cite{yin2023connectivity} focused on maximizing the user's connectivity via joint sub-carrier association and power allocation. However, the authors relied on traditional optimization approaches and proposed a heuristic solution, which is suboptimal and requires reformulation under the system condition variation. The study in~\cite{lien2017efficient,blanquez2016eolla,lai2015path} considered the role of feedback in solving the problem of resource allocation for mMTC and URLLC. However, feedback schemes require continuous channel estimation information exchanges, which causes high overhead and delay. The work in~\cite{park2020extreme,tang2024learn} adopted machine learning (ML) methods to support both mMTC and URLLC. Unfortunately, user equipment have limited computing power and energy efficiency, which makes applying ML models locally at UEs not applicable. Therefore, dynamic scheduling is a promising solution to support IIoT UEs with various intents, where the centralized scheduler at the gNB exploits the variation in the channel quality and traffic to efficiently allocate the available RBs without any collision. It also reduces signaling overhead and delay. However, traditional dynamic scheduling strategies rely on optimization frameworks, which require new formulations and solutions whenever any variation happens in the system.

\subsection{Contribution}

In this study, we propose a machine-learning framework for a centralized scheduler to grant access to IIoT UEs with various intents (i.e., URLLC requirements) and random traffic arrivals. The motivation behind using ML-based schedulers is the lack of theoretical performance guarantees, as is the case with heuristic schedulers. Besides, training an ML-based scheduler might uncover scheduling policies that heuristic practitioners might have overlooked. Another motivation is the data availability, where the scheduling of radio resources happens thousands of times per second, so there's plenty of data to train the ML models. However, one of the main challenges for applying it in a centralized scheduler while supporting IIoT UEs is the huge state and action space due to the number of user equipment and resource blocks in the system. Therefore, in this work, we address this challenge by proposing a graph-based reduction scheme to the state and action space that utilizes the nature of the problem. Our main contribution can be summarized as follows:

\begin{itemize}

\item We investigated the problem of supporting IIoT UEs with various intents in a computational offloading system with random task arrival traffic.
\item We proposed a machine-learning framework for a centralized scheduler to grant access to IIoT UEs with various intents. The framework utilizes reinforcement learning to adapt to the dynamic changes in the computational offloading system. A graph-based reduction scheme to the state and action space of the RL framework is proposed to help the centralized scheduler converge faster and learn an efficient scheduling strategy. 
\item Simulation results demonstrate the effectiveness of the proposed DRL-based dynamic scheduler in maximizing the number of successful task completion. We compared the proposed scheduling strategy with traditional scheduling approaches such as semi-static, round-robin, and heuristic. We also compared it with contention-free and contention-based approaches. The results show the significant performance of the proposed DRL-based scheduling strategy compared to these approaches. In addition, the proposed state and action reduction technique shows significant improvement in convergence compared to the proposed approach without reduction. 

\end{itemize}

The rest of the paper is organized as follows. In Section~\ref{system}, we state our system model. In Section~\ref{Problem}, we formulate the problem using a reinforcement learning framework and introduce our proposed solution. In Section~\ref{MAPPOsim}, we explain the PPO reinforcement learning algorithm. Section~\ref{simulation} provides our simulation model and results. Finally, we conclude the paper in Section~\ref{conclusion}.

\section{System Model}\label{system}

\subsection{Network Model}

We consider a multi-user uplink system consisting of a base station (gNB) and $N$ IIoT UEs indexed by~$\mathcal{N} =\{1,2,\dots,N\}$ as shown in Fig.~\ref{systemmodel}. The base station (BS) has~$M$ uplink orthogonal shared channels indexed by~$\mathcal{M} = \{1,2,\dots,M\}$ each with bandwidth~$W$ MHz and a multi-core central processing unit (CPU) with computation speed $F_{\mathrm{max}}$ cycles per second. The BS has a scheduler (i.e., centralized coordinator) that assigns the shared channels to IIoT UEs and allocates the computation resources equally among the scheduled UEs to allow parallel computation~{\footnote{The computation resources are equally allocated among scheduled IIoT UEs for the sake of simplicity and making the main problem under investigation, which is the uplink channel allocation tractable.}. Each IIoT UE~$n$ has a computation task queue with capacity~$K$ operates in a first-in-first-out (FIFO) manner, where the computation tasks are indexed by~$\mathcal{K} = \{1,2,\dots,K\}$. The computation tasks arrive at each IIoT UE according to the Poisson process with arrival rate~$\lambda = p_k \times T$, where~$p_k$ is the
task arrival probability and $T$ is the communication time period. Each computation task has parameters~$(A_k, C_k, \tau_k)$, where~$A_k$ is the computation task size in bits, $C_k$ is the number of required CPU cycles per bit and~$\tau_k$ is the task deadline constraint. Each IIoT UE offloads each computation task in the buffer to the BS with a fixed power level $p_n$ depending on the communication resources assigned by the BS. Each IIoT UE expresses his intent (i.e. required quality of service (QoS)) to the network through a graphical user interface (GUI). The GUI consists of a drop-down menu, which associates a set of values for each QoS attribute. We consider the transmission of computation tasks to be finished when each IIoT UE’s computation queue is empty. 

\begin{figure} 
\centering
\includegraphics[width= 3.2 in, height = 3.5 in]{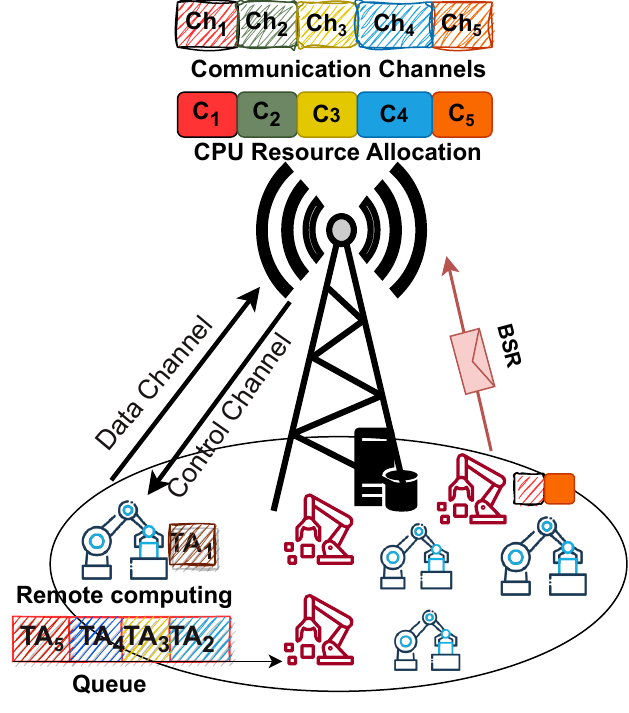}
\caption{System Model.}
\label{systemmodel}
\end{figure}

\subsection{Communication Model}

The BS and IIoT UEs communicate over data and control channels. The data are sent over data channels while control information is exchanged over control channels. We assume that the uplink and downlink control channels are dedicated and error-free. The uplink data channel gain between an IIoT UE~$n$ and the gNB on subcarrier~$m$ $h_{n,m} = g_{n,m} \sqrt{\beta_m}$ consists of a small-scale fading~$g_{m,k}$ and a large-scale fading~$\sqrt{\beta_m}$. $\boldsymbol{z} \sim\mathcal{CN}(0,\sigma^2)$ represents an additive white Gaussian noise (AWGN) vector, where $\sigma^2 = \aleph_0 W$ and $\aleph_0/2$ is the noise power spectral density. At each time step~$t$, each IIoT UE sends the buffer status report (BSR) and requested QoS on the control channel and offloads its computation task if the scheduler assigns an uplink data channel to IIoT UEs. We consider a discrete system model that adopts an orthogonal frequency division multiple access (OFDMA) transmission scheme. At each time slot, the BS allocates the uplink data channels to IIoT UEs to offload their computation tasks. The received signal at the BS on subcarrier~$m$ is expressed as 
\begin{equation}
y_m = \sum_{n \in \mathcal{N}_m} h_{n,m} s_{n,m} + z_m
\end{equation}
where the set~$\mathcal{N}_m$ denotes the subset of IIoT UEs transmitted data on subcarrier~$m$, $s_{n,m}$ denotes the transmitted signal of the $n$-th IIoT UE over the $m$-th subcarrier and $z_m$ is the noise power at the BS. The BS adopts a non-orthogonal multiple access (NOMA) transmission mode at each subcarrier to accommodate multiple connectivity, where each IIoT UE is allocated to at most one subcarrier at each time slot and offloads one task~\footnote{{Note that IoT devices are numerous and have limited transmit power, so they are more
likely to achieve a connection by focusing the power in one single subcarrier.}}. Based on~\cite{meredith2015study}, in NOMA mode at most two IIoT UEs are allocated to each subcarrier to reduce error propagation and decoding complexity of successive interference cancellation (SIC). Assuming that two IIoT UEs indexed by~$\{n,n'\}$ allocated to a subcarrier~$m$, where~$|h_{n,m}|^2 \geq |h_{n',m}|^2$, the achievable data rates are given by
\begin{equation}
R_{n,m} = W \log(1+ \frac{|h_{n,m}|^2p_{n}}{\sigma^2 + \sum_{i \in \mathcal{N}_m \setminus n}|h_{i,m}|^2p_{i}})     \; \; \textit{bps}
\end{equation}

\begin{equation}
R_{n',m} = W \log(1+ \frac{|h_{n',m}|^2p_{n}}{\sigma^2})     \; \; \textit{bps}
\end{equation}

\subsection{Computation Model}

We adopt a computation model based on the advanced dynamic voltage and frequency scaling (DVFS) technique at the BS~\cite{you2016energy}. At a time slot~$t$, if the IIoT UE~$n$ is allocated a channel and its computation task queue is non-empty, its computation task is executed remotely at the BS. In this case, the total delay for remote computation consists of the uplink transmission time, the task execution time at the BS, and the downlink transmission time. Since the BS has high transmission power and the computation results size is usually small compared to offloaded data~{\footnote{Note that we consider a smart logistics use case, where mobile robots read the RFID of goods and get information about the goods like size, type etc. then offload the information to the BS to compute the best path for goods from a vehicle to shelves and vice versa.}}, we ignore the downlink transmission time. Thus, the remote computation time consisting of the uplink transmission time~$t^{\mathrm{up}}$ and remote execution time~$t^{\mathrm{exe}}$ is given by 
\begin{equation}
t^r_{n,k} = t^{\mathrm{up}} + t^{\mathrm{exe}} = \frac{A_{n,k}}{R_{n,m}} + \frac{A_{n,k} \times C_{n,k}}{f_{k,m}}    
\end{equation}
where~$f_{k,m}$ is the computation resources allocated to task~$k$ on subcarrier~$m$ and~$R_{n,m}$ is the uplink rate of IIoT UE~$n$ on subcarrier~$m$. Each IIoT UE deletes the computation task from the computation queue if it is successfully computed remotely; otherwise, the task remains until served.

\subsection{Performance Metrics}

The URLLC is guaranteed to an IIoT UE~$n$ in a time slot~$t$ if the requested latency is satisfied and reliability is achieved. Reliability is defined by achieving an outage probability less than or equal to a certain threshold $p_{\mathrm{out}} \leq \epsilon$, where~$p_{\mathrm{out}} = \mathrm{Pr}\{R_{n,m} \leq r_{\mathrm{th}}\}$, where $r_{th}$ is a transmission rate threshold. {\em In this study, our objective is to maximize the number of computation tasks while achieving the required URLLC}.

\section{Problem Formulation and Proposed Solution}~\label{Problem}

We formulate the URLLC task scheduling problem as a reinforcement learning problem due to the dynamic variation in the wireless system and NP-Hardness of the problem~\cite{destounis2018scheduling,destounis2019complexity}. We aim to maximize the number of computation tasks that can be executed within the deadline constraint while achieving the required reliability for IIoT UEs. 
\begin{itemize}
\item \textbf{State Space:} the BS state consists of a tuple containing the recent~$l$ observations~$(o^b_t,\dots,o^b_{t-l})$, which include the buffer status of all IIoT UEs, their deadline constraints, and the channel gain matrix.
\item \textbf{Action Space:} the BS takes an environment action, which is an allocation of the uplink communication channels to a subset of IIoT UEs denoted by~$\mathcal{N}_s$. Note that the BS environment action space contains all possible channel allocations to each pair of IIoT UEs. For example, if we consider $N = 4$ users and $M=2$ channels, the action space consists of~ $90$ possible actions, where we consider all possible combinations of IIoT UEs including no IIoT UEs allocation expressed as~$$ C = \binom{|\mathcal{N}| + 1}{|\mathcal{N}_s|}.$$ Then, we consider all possible permutations of the subset of IIoT UEs on the communication channels. Thus, the total number of possible actions is calculated as $$\textit{no. of actions} = {C}P_{M} = \dfrac{C!}{(C - M)!}.$$
The calculation is as follows: $ C = \binom{|\mathcal{N}| + 1}{|\mathcal{N}_s|} = \binom{5}{2} = 10$. Then, $\textit{no. of actions} = \dfrac{10!}{(10 - 2)!} = 90.$ 
\item \textbf{Reward function:} the utility function is defined to maximize the number of successfully executed tasks. Therefore, at each time step, the base station reward~$R_b(t)$ is defined as
$$R_b(t) = \sum_{n \in \mathcal{N}_s} R_n(t).$$ the sum of the rewards from the subset of IIoT UEs~$\mathcal{N}_s$ allocated to the communication channels. Each IIoT UE reward~$R_n(t)$ is defined as
$$R_n(t) = \begin{cases}
+\rho & \text{ if the task execution is successful}\\
-\rho & \text{otherwise} \\ 
\end{cases}$$
where the reward is $+\rho$ if the IIoT UE task is successfully executed through the allocated channel,~$-\rho$ otherwise.
\end{itemize}

To solve the URLLC task scheduling problem, there is a main challenge that needs to be tackled, which is the huge state space at the BS. The huge state space is caused by the large observation due to the number of IIoT UEs and resource blocks.  The huge state space causes slow learning and poor generalization to new unseen scenarios. To tackle the huge state space challenge and allow the centralized scheduler to learn an efficient channel assignment protocol, we propose a framework that relies on pre-processing (i.e. reducing) the state space before passing it to the RL policy network. The following concept from graph theory is needed~\cite{lovasz2009matching}:

\begin{mydef}
A weighted hypergraph $\mathcal{H=(V,E)}$ is a set of vertices $\mathcal{V}$ and a collection $\mathcal{E}$ of subsets of the vertex set called hyperedges with a weight $w_e$ assigned to each $e \in \mathcal{E}$. A hypergraph is called $k$-uniform if each hyperedge has exactly $k$ vertices. The degree of a vertex is the number of hyperedges to which it belongs.
A matching on $\mathcal{H}$ is a subset of $\mathcal{E}$, in which every two hyperedges are disjoint.  
\end{mydef}

The framework works as follows: the centralized scheduler constructs a 3-uniform hypergraph~$\mathcal{H=(V,E)}$, where the set of vertices~$\mathcal{V} = \mathcal{N} \cup \mathcal{M}$ is the union of the set of IIoT UEs and the set of uplink communication channels. The set~$\mathcal{E}$ contains a collection of hyperedges such that a hyperedge~$e \in \mathcal{E}$ consists of $\{n,n',m\} \subseteq \mathcal{V}$, where $n,n' \in \mathcal{N}$ (with $n \neq n'$) and $m \in \mathcal{M}$. The IIoT UEs are divided into two different groups far and nearby based on their distance to the BS. The hyperedges are constructed such that the IIoT UEs belonging to a hyperedge~$e \in \mathcal{E}$ are chosen from two different groups. Afterward, the constructed hypergraph is passed to the RL policy network, where the action space consists of all possible matching of size equal to the number of uplink communication channels on the constructed hypergraph. 

The intuition behind our proposed framework comes from realizing the nature of our problem instead of blindly listing all possible actions as input to the learning algorithm. It is well-known that nearby IIoT UEs do not lead to successful SIC in NOMA, thus we divide the IIoT UEs into two groups and allow pairing between IIoT UEs from different groups only. Moreover, the number of IIoT UEs in the system and limited communication resources give a high likelihood of finding a high number of IIoT UEs having non-empty buffers justifying considering only matching of size equal to the number of uplink communication channels. The following simple example illustrates how the proposed framework can reduce the state space and help the centralized scheduler learn an efficient channel assignment protocol. 
\begin{myexam}\label{Example1}
Consider four IIoT UEs indexed by~$\mathcal{N} =\{1,2,3,4\}$ with non-empty buffers and two communication channels indexed by~$\mathcal{M} = \{1,2\}$. The action space before reduction consists of~$90$ possible actions as explained above. To reduce the action space, the centralized scheduler divides the IIoT UEs into a set of far IIoT UE~$\mathcal{N}_f = \{1,2\}$ and a set of near IIoT UE~$\mathcal{N}_r = \{3,4\}$ based on their channel gains to the BS. Then, pairing between IIoT UEs is allowed only between IIoT UEs from different sets. Since we aim to maximize the number of successful computation tasks, the centralized scheduler considers only possible permutations of size equal to twice the number of communication channels from the possible paired IIoT UEs. As a result, the action space after this reduction consists of ~$4$ actions only listed in the following set 
$$\begin{array}{l}\biggl\{\Bigl\{\{1, 3,\bold{1}\}, \{2, 4,\bold{2}\}\Bigl\}, \Bigl\{\{2, 4,\bold{1}\},\{1, 3,\bold{2}\}\Bigl\},\\
\Bigl\{\{1, 4,\bold{1}\},\{ 2, 3,\bold{2}\}\Bigl\},\Bigl\{\{2, 3,\bold{1}\},\{1, 4,\bold{2}\}\Bigl\}\biggl\}
\end{array}$$
As we can see, we can scale up easily with our proposed framework due to its ability to significantly reduce the action space to the most effective actions that can maximize our utility function. 
\end{myexam}

\section{Proximal Policy Optimization (PPO)-based  scheduling approach}~\label{MAPPOsim}

The RL agent (i.e., gNB scheduler) needs to learn a good policy to schedule the IIoT UEs. The RL agent interacts with the environment and explores different states and actions to evaluate and optimize
the policy iteratively. We adopt the proximal policy optimization (PPO) algorithm~\cite{schulman2017proximal,yu2021surprising}, which is an on-policy policy gradient algorithm as it supports large action spaces better than value-based RL methods. The actor-critic algorithm~\cite{haarnoja2018soft} is adopted along with the generalized advantage estimation (GAE)~\cite{schulman2015high}. The actor-critic network architecture is constructed, where the policies and the value function(s) are distinct and consist of four-layer D2RL~\cite{sinha2020d2rl}. The D2RL relies on using deeper MLP layers along with dense connections, where the original input features (i.e. state or state-action pairs) are 
passed to each hidden layer concatenated with the previous hidden layer output except the last hidden layer since it is a direct linear transformation. This structure avoids decreasing the mutual information between the input and output due to the non-linear transformations used in deep learning, which is called data processing inequality (DPI). The D2RL has shown significant improvement in the performance and sample efficiency compared to the vanilla two-layer MLP network architecture. 

To update the weights of the neural networks, the network keeps track of a fixed-length trajectory of memories (i.e., states, actions, rewards). Then, the neural network is updated using stochastic gradient ascent via mini-batches, where ~$ 4$ epochs of updates are done on each mini-batch. The actor update rule optimizes the surrogate-clipped objective function
\begin{equation*}
L^{\mathrm{CLIP}}(\theta)  = \hat{\mathbb{E}}_t\big[\min(r_t(\theta)\hat{\boldsymbol{A}_t},\mathrm{clip}(r_t(\theta), 1 - \epsilon, 1 + \epsilon)\hat{\boldsymbol{A}_t}\big],
\end{equation*}
where~$\epsilon$ is a hyperparameter and~$r_t(\theta) = \frac{\pi_{\mathrm{\theta}}(a_t | s_t)}{\pi_{\mathrm{\theta}_{\mathrm{old}}}(a_t | s_t)}$ is the ratio between the new and old policies. The advantage at each time step is calculated by 
\begin{equation*}
\hat{\boldsymbol{A}_t} = \delta_t + (\gamma \lambda) \delta_{t +1} + \dots + \dots +  (\gamma \lambda)^{T -t +1} \delta_{T - 1}, 
\end{equation*}
where~$\delta_t = r_t + \gamma V^{\phi}(s_{t+1}) - V^{\phi}(s_t)$ counts the benefits of the new state over the old state. The value update rule is done to optimize the advantage function 
\begin{equation*}
J(\phi) = [r_t + \gamma V^{\phi}(s_{t+1}) - V^{\phi}(s_t)]^2.    
\end{equation*}
The actor decides the next action based on the current state, while the critic evaluates the states. The PPO-based algorithm for task scheduling is described in Algorithm~\ref{IPPO}. {The computational complexity of PPO during inference is~$\mathcal{O}(S)$, where $S$ is the number of time steps in an episode since it depends on the learned policy. During training, the computational complexity is~$\mathcal{O}( B \times E \times S)$, where $B$ is the number of batches, $E$ is the number of episodes in each batch, $S$ is the number of time steps in an episode \cite{qi2024efficient}.}

\begin{algorithm}[t]
\SetKwInOut{Input}{Input}
\SetKwInOut{Output}{Output}
\begin{algorithmic}[1]
\STATE Initialize the policy network~$\pi_{\theta}(a|s)$ with~$\theta$.
\STATE Initialize the critic network~$V_{\phi}(s)$ with~$\phi$.
\STATE Initialize the reply buffer~$B$.
\FOR{$t$ in $\{1,2,\dots,T\}$}
\STATE Take actions according to the policy~$\pi_{\theta}$ to generate a trajectory~$\{(s_1,a_1,r_1),\dots,(s_t,a_t,r_t)\}$ and store them to~$B$;
\IF{$t\%B == 0$}
\STATE Calculate the reward~$r(t)$ and advantage estimate~$\hat{\mathbf{A}}(t)$;
\STATE Calculate~$\nabla_{\theta} L$ and ~$\nabla_{\phi} J$;
\STATE Update~$\theta,\phi$;
\STATE Clear the reply buffer~$B$;
\ENDIF
\ENDFOR
\RETURN Scheduled Tasks.
\end{algorithmic}
\caption{PPO-based Task Scheduling Algorithm}~\label{IPPO}
\end{algorithm}

\section{Simulation Model and Results}\label{simulation}

We present our simulation model and evaluate the performance of our proposed framework in this section. The proposed framework is evaluated in a warehousing logistic area of~$100 \times 100$~m$^2$ containing a BS located at the center and~$30$ IIoT UEs randomly distributed according to a homogeneous Poisson point process over the area. 
The signal attenuation from the BS to IIoT UEs is based on the 3GPP standard \cite{3GPP} with a distance-based path loss model~$128.1 + 37.6 \log_{10} d$, where~$d$ is the distance in kilometers and Rayleigh fading with unit variance. {Note that in our study, we assume an indoor smart logistic use case where mobility is very limited to small areas (i.e., low mobility) and resources can still be supported without the need for high adjustment to mobility patterns as in medium and high mobility scenarios.} The simulation parameters for the system are listed in Table~\ref{table:1}, and the main parameter settings for the learning algorithm are given in Table~\ref{table:2}. The policy and value functions are implemented separately and optimized by Adam optimizer~\cite{kingma2014adam}. We use the same neural network architecture for both the actor and the critic. The results are averaged over~$8$ random seeds, where the location of IIoT UEs, intents, and traffic arrival probability changed between episodes. The performance is evaluated over the training process, where the solid lines represent the average performance in the evaluation episodes during the training and the shaded regions show the~$95\%$ confidence interval (CI). We compare the proposed framework with the following benchmarks:~\footnote{Although the proportional fair (PF) scheduler is a very common scheduler, we did not include it in our comparison since it optimizes the throughput allocated to a user at a time slot compared to average throughput that the user has received over time and our objective in this work is maximizing the number of successful computation tasks under URLLC constraints; thus, implementing the PF scheduler will not bring a fair comparison with our proposed scheme.}

\begin{table}
\caption{Simulation Parameters}
\centering
\begin{tabular}{|c|c|}
\hline
\textbf{Parameters} & \textbf{Values}\\
\hline
No. of sub-carriers $(M)$ & $3$ \\
\hline
Carrier frequency band & $410–7125$~MHz\\
\hline
Subcarrier spacing & $15$~kHz \\
\hline
System bandwidth $(W)$ & $30$~MHz\\
\hline
No. of IIoT UEs $(N)$ & $30$ \\
\hline
Distance-dependent Path loss & $128.1 + 37.6 \log_{10} d,$ dB\\
\hline
Channel Fading Model & Rayleigh fading \\
\hline
Tasks Size $(A_k)$ & $100-500$ bits\\
\hline
Tasks Computation Requirement $(C_k)$ & $1 \times 10^2- 2 \times 10^4$\\
\hline
Tasks Delay Tolerance~$(\tau_k)$ &$1 - 5$~millisecond\\
\hline
IIoT UEs Reliability Requirement $(\epsilon)$ & $10^{-3}$ \\
\hline
Noise Power Spectral Density & -174 dBm/Hz\\
\hline
BS Computation Capacity~$(F_{\mathrm{max}})$ & $120$~GHz  \\
\hline
IIoT UE Queue Capacity~$(K)$ & $50$\\
\hline
IIoT UE transmit power level $(p_n)$ & $0.08$~Watt\\
\hline
Probability of Task Arrival~$(p_k)$ & $0.80$\\
\hline
Duration of episode & 25\\
\hline
\end{tabular}
\label{table:1}
\end{table}

\begin{table}
\caption{PPO Hyperparameters}
\centering
\begin{tabular}{|c|c|c|c|}
\hline
$\textbf{Hyperparameter}$ & $\textbf{Values}$ & $\textbf{Hyperparameter}$ & $\textbf{Values}$\\
\hline
No. of episodes & 6000 & Entropy coeff.~$(c2)$ & $0.01$  \\
\hline
Minibatch size & $32$ & Discount factor~$(\gamma)$& $0.99$\\
\hline
GAE parameter~$(\lambda)$ & $0.95$ & Clipping parameter~$(\epsilon)$ & $0.2$\\
\hline
Optimizer & Adam & Optimizer epsilon & $10^{-5}$\\
\hline
Activation function & Relu & No. of hidden layers & 1\\
\hline
Actor Size & $256$ & Critic Size & $512$ \\
\hline
Actor learning rate & $10^{-2}$ & Critic learning rate& $10^{-4}$\\
\hline
\end{tabular}
\label{table:2}
\end{table}

\begin{itemize}
\item \textbf{Contention-based:} Each IIoT UE transmits with a certain probability~$p_t$ if the computation queue is not empty and randomly accesses the uplink channels.

\item \textbf{Contention-free:} The BS controls and schedules the transmission over the uplink data channels. Each IIoT UE sends a scheduling request if its computation queue is not empty and offloads the task if the BS sends a scheduling grant. The IIoT UE deletes the task from the queue if it satisfies the latency and reliability constraints and an ACK is received from the BS. 
If the IIoT UE made a successful task computation and sent a scheduling request simultaneously, the BS will send an ACK and ignore the scheduling request.

\item \textbf{Semi-static scheduling}: The BS pre-allocates the resources to IIoT UEs. Then, whenever there are computation tasks in their queues, the IIoT UEs offload them to the BS~\cite{dahlman20205g}.

\item \textbf{Round-robin scheduling}: IIoT UEs take turns to get access to the resource blocks~\cite{arpaci2014chapter}.

\item \textbf{Heuristic scheduling}: Algorithm~\ref{greedy} finds a suboptimal solution to the maximum cardinality matching in the constructed hypergraph with low computational complexity as follows. For each vertex~$v \in \mathcal{M}$, we find the subset of hyperedges that contains the vertex $v$, denoted by $\mathcal{E}_v$. If $\mathcal{E}_v$ is not empty, we find the hyperedge~$e \in \mathcal{E}_v$ that has the maximum weight if there is a tie, we choose one randomly. The weight represents the number of successfully computed tasks of the IIoT UEs in the hyperedge. Then, we add it to the matching set denoted by  $\mathcal{E}'$. Afterward, the hyperedge~$e$ and all the hyperedges intersecting with it are removed from the set of hyperedges $\mathcal{E}$. The same procedure is repeated for each $v$ in $\mathcal{M}$.
\end{itemize}

\begin{algorithm}[t]
\SetAlgoLined
\SetKwInOut{Input}{Input}\SetKwInOut{Output}{Output}
\Input{$3$-uniform hypergraph $\mathcal{H=(V,E)}$, with hyperedge weight $w'_e$ for each $e \in \mathcal{E}$;}
\Output{Maximum Weighted Matching $\mathcal{E}'$;}
\begin{algorithmic}[1]
\FOR{each $v \in \mathcal{M}$ }
\STATE Find the subset of hyperedges $\mathcal{E}_v \subseteq \mathcal{E}$ that contains \\ the vertex~$v$;
\IF{$\mathcal{E}_v$ is not empty}
\STATE Choose~$e \in \mathcal{E}_v$ that has the maximum weight and \\ add it to $\mathcal{E}'$;
\STATE Remove $e$ and all the hyperedges that intersect \\ with it from $\mathcal{E}$;
\ENDIF
\ENDFOR
\RETURN $\mathcal{E}'$;
\end{algorithmic}
\caption{Heuristic Algorithm for Hypergraph Maximum Weighted Matching}
\label{greedy}
\end{algorithm}

Fig.~\ref{Fig1} shows the normalized successful computed tasks satisfying the latency and reliability constraints versus the number of training episodes. As we can observe, our proposed DRL-based scheduling scheme with reduction outperforms the same scheme without reduction in maximizing the number of successfully computed tasks and fast convergence. This is due to the reduction in state and action spaces, which assists the RL algorithm in learning and converging faster. Moreover, we can see that the DRL-based proposed scheme beats the traditional scheduling approaches round-robin, semi-static, and heuristic. The main reason is that the proposed approach takes into consideration the dynamic changes in the traffic arrival, various requested QoS, and channel variation, which efficiently utilizes the available resources compared to traditional schemes. It also outperforms the random access scheme contention-based due to the centralized channel allocation, which reduces collision. It also performs better than the contention-free scheme due to the reduction in the signaling control messages overhead.

\begin{figure}
    \centering
    \includegraphics[width = 3.5 in, height = 3in]{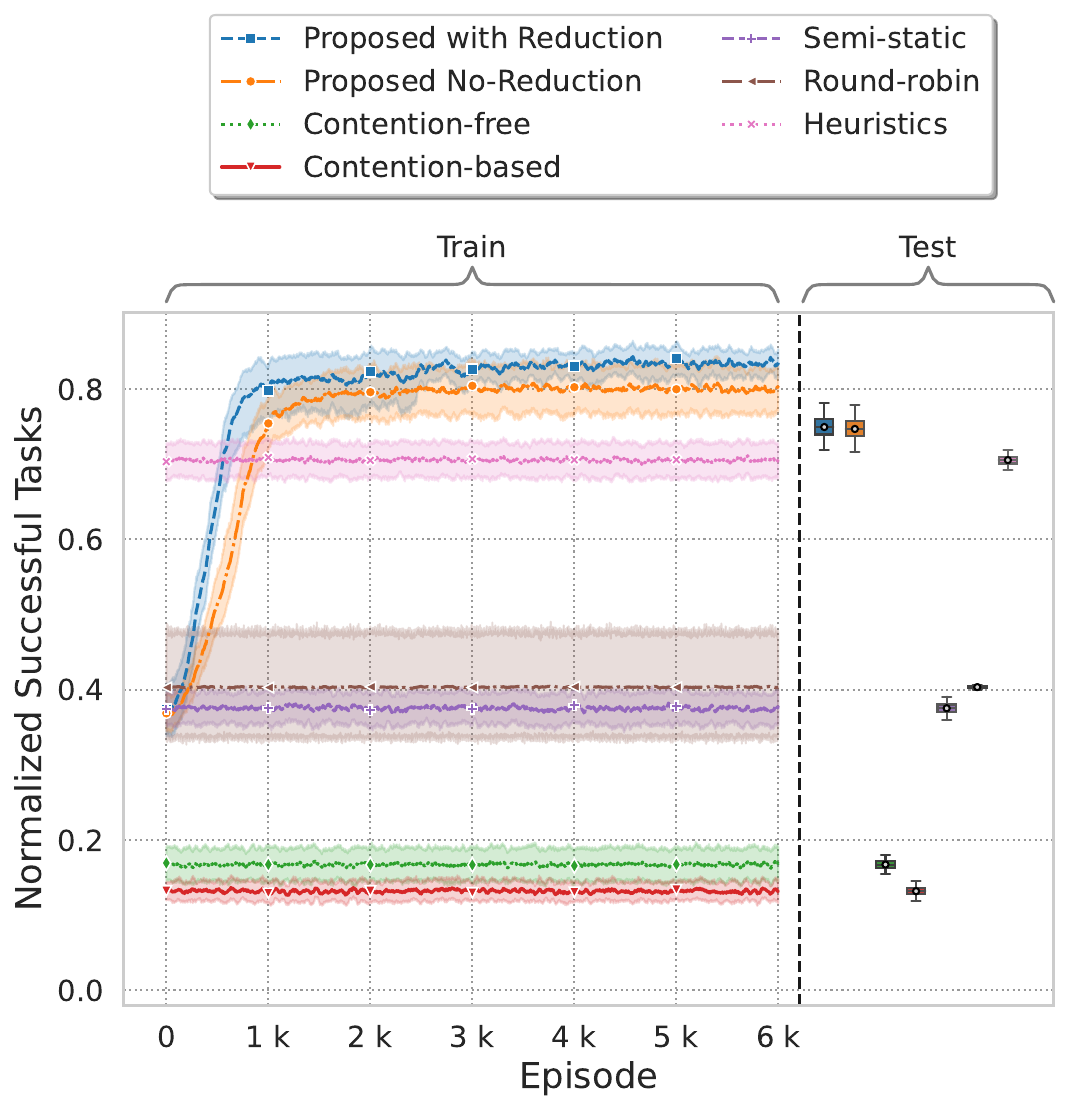}
    \caption{Normalized successful computed tasks versus training episodes.}
   \label{Fig1}
\end{figure}

Fig.~\ref{Fig2} shows the normalized failed tasks due to unsuccessful computation versus training episodes. As we can see, the DRL-based proposed scheme reduces the number of failed tasks compared to traditional scheduling, contention-based, and contention-free schemes. The reason is that the learned scheduling policy adapts to the dynamic changes in the arrival traffic, various requested QoS, and channel conditions. Fig.~\ref{Fig3} demonstrates the system goodput rate versus training episodes. As we can notice, the DRL-based proposed scheme has a high goodput rate compared to traditional scheduling, contention-based, and contention-free schemes due to the centralized channel allocation and adaptation to the changes in the traffic arrival and channel conditions.

\begin{figure}
    \centering
    \includegraphics[width = 3.5 in, height = 3 in]{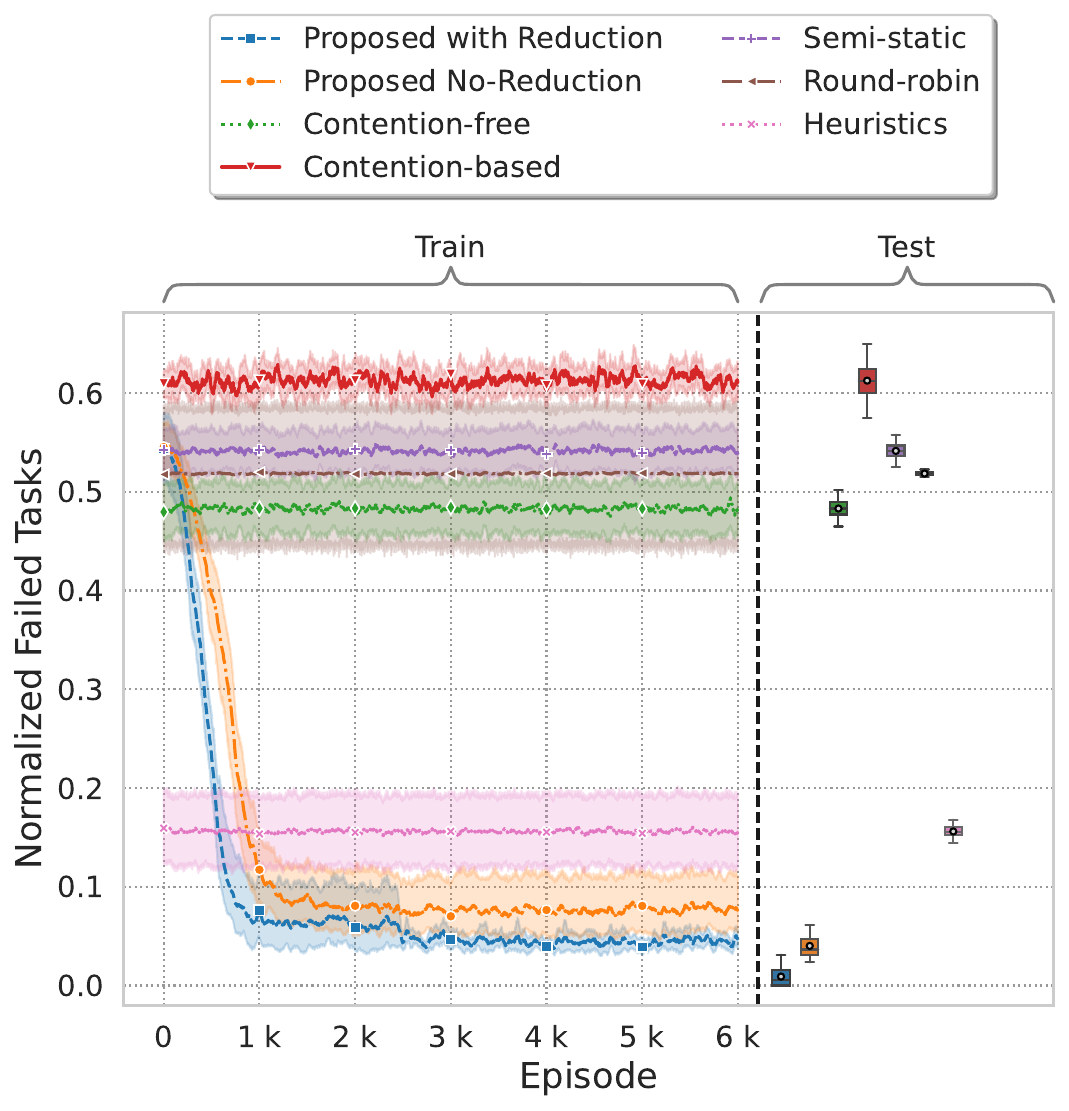}
    \caption{Normalized failed tasks versus training episodes.}
   \label{Fig2}
\end{figure}

\begin{figure}
    \centering
    \includegraphics[width = 3.5 in, height = 3 in]{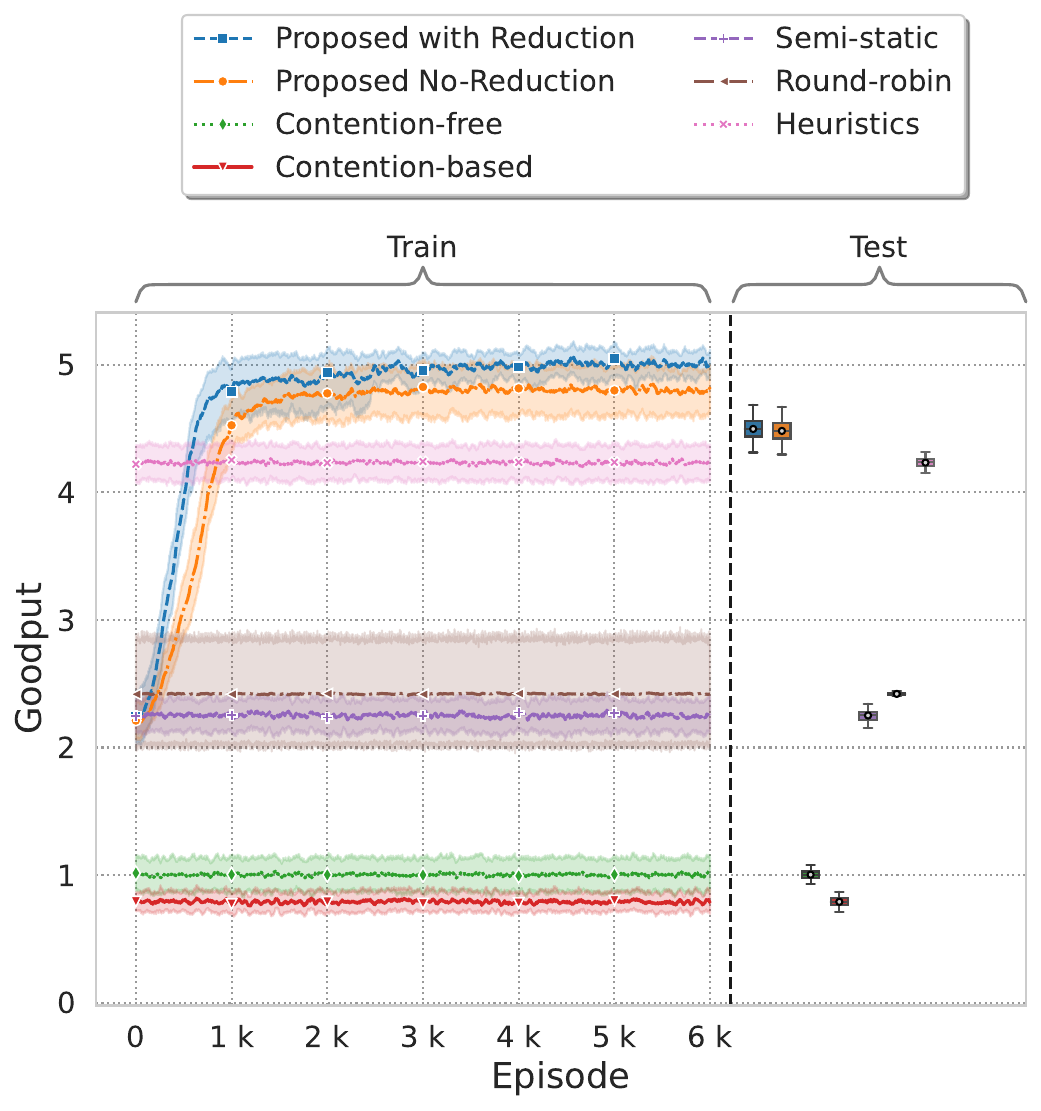}
    \caption{Goodput versus training episodes.}
   \label{Fig3}
\end{figure}

We investigated the channel collision rate for different schemes and we found that DRL-based proposed scheme, traditional scheduling schemes, and contention-free have zero collision rates due to the centralized channel allocation at the base station compared to the random access scheme contention-based. Fig.~\ref{Fig6} shows the performance of the PPO algorithm under different network architectures, where single layer and D2RL are compared. As we can observe, the two architectures give close performance during training. However, D2RL shows better performance during testing because of its depth, ability to learn hierarchical features and use of dense connections for training stability. Fig.~\ref{Fig7} demonstrates the performance of the PPO algorithm under different actor and critic model sizes. As we can see, the small model size gives a better performance during the training and testing phases. The reason is that smaller networks have fewer parameters, which makes them less prone to overfitting, generalize better to unseen data, and focus on learning the essential patterns.




\begin{figure}
    \centering
    \includegraphics[width = 3.5 in, height = 3 in]{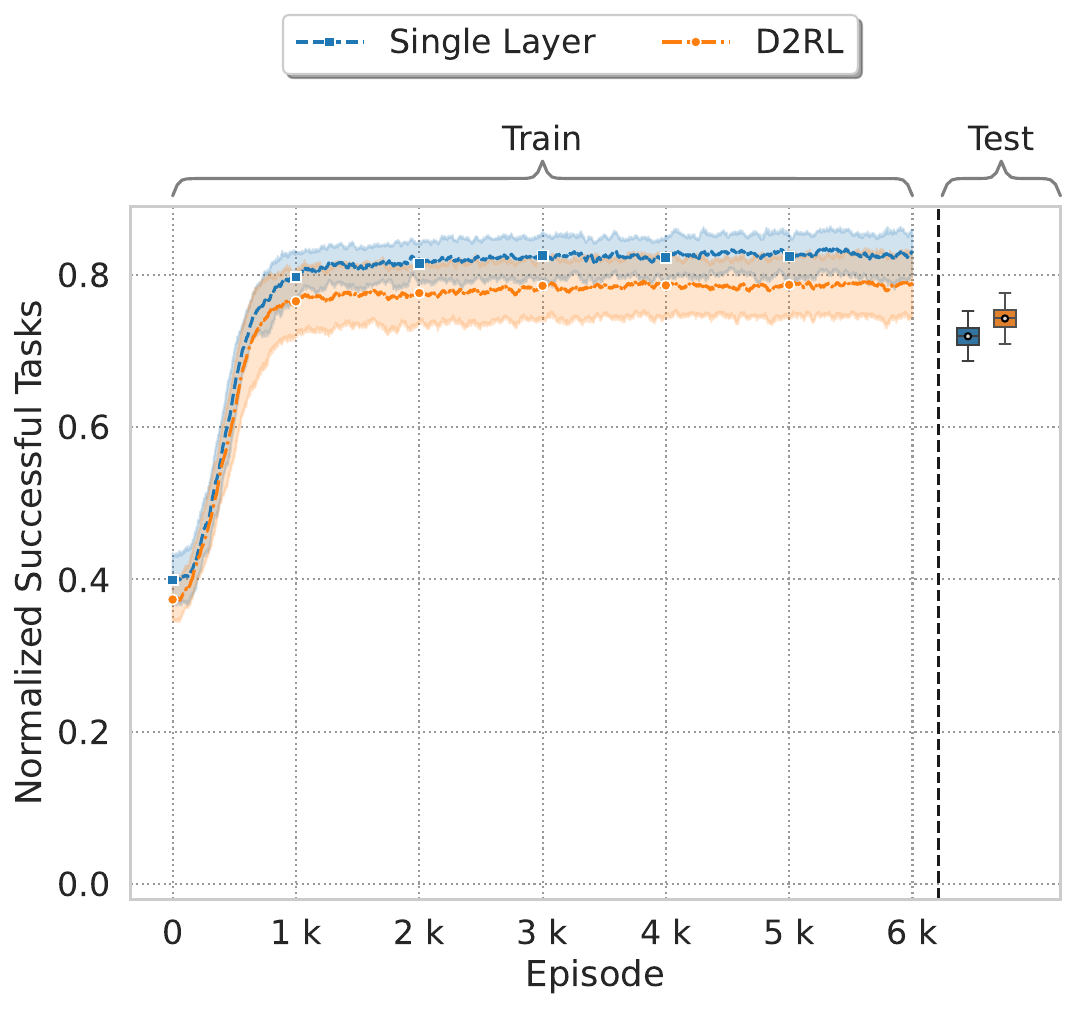}
    \caption{Comparison between single layer and D2RL network architectures.}
   \label{Fig6}
\end{figure}

 \begin{figure}
    \centering
    \includegraphics[width = 3.5 in, height = 3 in]{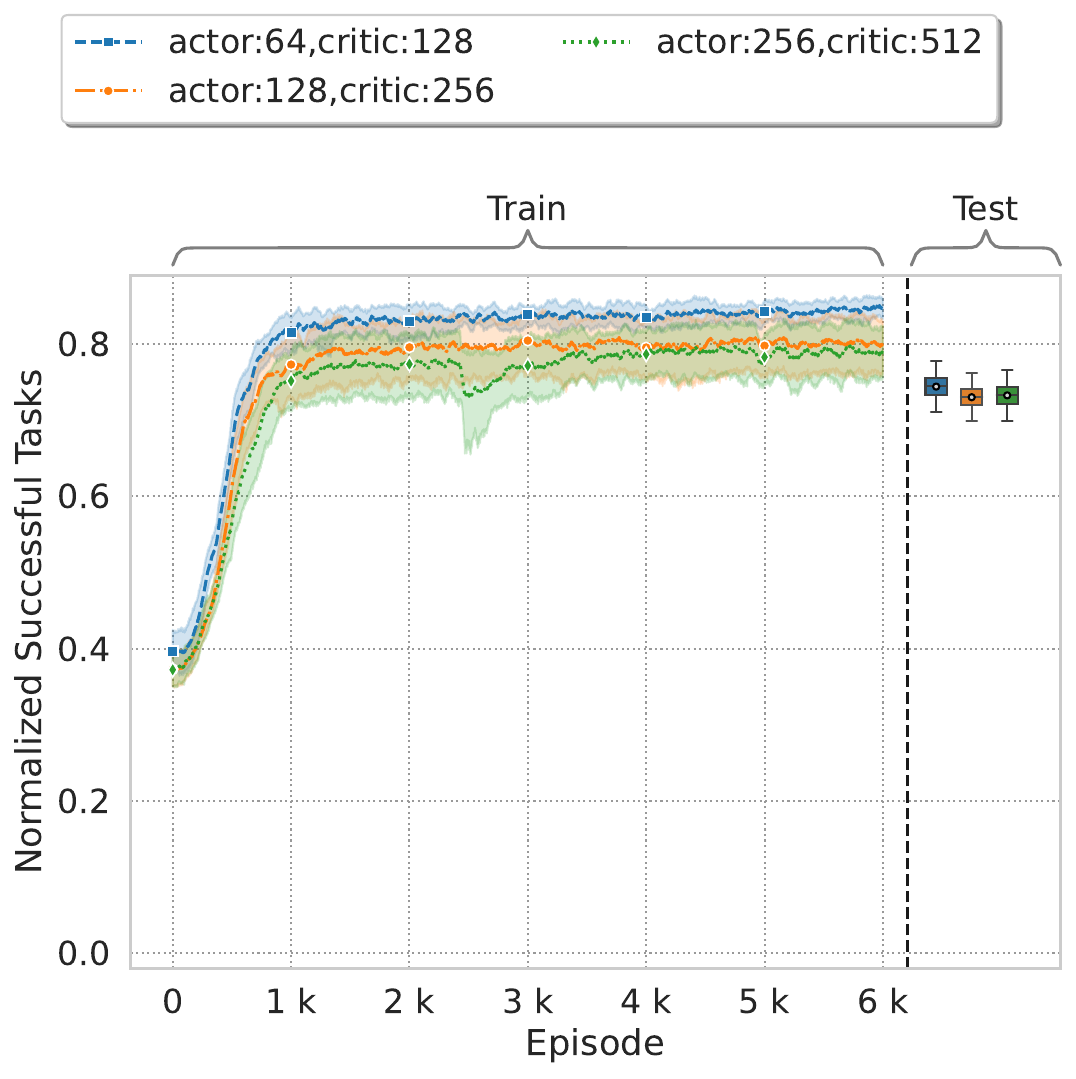}
    \caption{Comparison between a single layer model with different sizes.}
   \label{Fig7}
\end{figure}

\section{Conclusion}\label{conclusion}

A DRL-based centralized dynamic scheduler is proposed to support IIoT UEs with various intents (i.e., URLLC requirements). The proposed scheduler leverges RL to adapt to the dynamic changes in wireless communication channels and traffic arrival. Moreover, a graph-based reduction strategy is proposed to help the scheduler learn an efficient scheduling policy and converge faster. The numerical results show the outperformance of the proposed scheduler in maximizing the number of successfully computed tasks and the goodput rate compared to traditional scheduling schemes such as round-robin, semi-static, and heuristic. It also outperformed the random access scheme contention-based and the signaling-based scheme contention-free. 

\section*{ACKNOWLEDGMENT}

The work is funded by the European Union through projects 6G-INTENSE (G.A no. 101139266), CENTRIC (G.A no. 101096379). Views and opinions expressed are, however, those of the author(s) only and do not necessarily reflect those of the European Union. Neither the European Union nor the granting authority can be held responsible for them. 

\bibliographystyle{ieeetr}
\bibliography{References}

\begin{thebibliography}{10}

\bibitem{dahlman20205g}
E.~Dahlman, S.~Parkvall, and J.~Skold, {\em 5G NR: The next generation wireless access technology}.
\newblock Academic Press, 2020.

\bibitem{garcia2021tutorial}
M.~H.~C. Garcia, A.~Molina-Galan, M.~Boban, J.~Gozalvez, B.~Coll-Perales, T.~{\c{S}}ahin, and A.~Kousaridas, ``A tutorial on {5G NR V2X} communications,'' {\em IEEE Communications Surveys \& Tutorials}, vol.~23, no.~3, pp.~1972--2026, 2021.

\bibitem{leivadeas2022survey}
A.~Leivadeas and M.~Falkner, ``A survey on intent based networking,'' {\em IEEE Communications Surveys \& Tutorials}, 2022.

\bibitem{clemm2020intent}
A.~Clemm, L.~Ciavaglia, L.~Z. Granville, and J.~Tantsura, ``Intent-based networking-concepts and definitions,'' {\em IRTF draft work-in-progress}, 2020.

\bibitem{abbas2021network}
K.~Abbas, T.~A. Khan, M.~Afaq, J.~J.~D. Rivera, and W.-C. Song, ``Network data analytics function for {IBN}-based network slice lifecycle management,'' in {\em 2021 22nd Asia-Pacific Network Operations and Management Symposium (APNOMS)}, pp.~148--153, IEEE, 2021.

\bibitem{navarro2020survey}
J.~Navarro-Ortiz, P.~Romero-Diaz, S.~Sendra, P.~Ameigeiras, J.~J. Ramos-Munoz, and J.~M. Lopez-Soler, ``A survey on {5G} usage scenarios and traffic models,'' {\em IEEE Communications Surveys \& Tutorials}, vol.~22, no.~2, pp.~905--929, 2020.

\bibitem{pokhrel2020towards}
S.~R. Pokhrel, J.~Ding, J.~Park, O.-S. Park, and J.~Choi, ``Towards enabling critical {mMTC: A review of URLLC within mMTC},'' {\em IEEE Access}, vol.~8, pp.~131796--131813, 2020.

\bibitem{qiu2020edge}
T.~Qiu, J.~Chi, X.~Zhou, Z.~Ning, M.~Atiquzzaman, and D.~O. Wu, ``Edge computing in industrial internet of things: Architecture, advances and challenges,'' {\em IEEE Communications Surveys \& Tutorials}, vol.~22, no.~4, pp.~2462--2488, 2020.

\bibitem{mao2017survey}
Y.~Mao, C.~You, J.~Zhang, K.~Huang, and K.~B. Letaief, ``A survey on mobile edge computing: The communication perspective,'' {\em IEEE communications surveys \& tutorials}, vol.~19, no.~4, pp.~2322--2358, 2017.

\bibitem{chapman2014hspa}
T.~Chapman, E.~Larsson, P.~von Wrycza, E.~Dahlman, S.~Parkvall, and J.~Skold, {\em HSPA evolution: The fundamentals for mobile broadband}.
\newblock Academic Press, 2014.

\bibitem{guan2018service}
W.~Guan, X.~Wen, L.~Wang, Z.~Lu, and Y.~Shen, ``A service-oriented deployment policy of end-to-end network slicing based on complex network theory,'' {\em IEEE access}, vol.~6, pp.~19691--19701, 2018.

\bibitem{zhang2017network}
H.~Zhang, N.~Liu, X.~Chu, K.~Long, A.-H. Aghvami, and V.~C. Leung, ``Network slicing based {5G} and future mobile networks: mobility, resource management, and challenges,'' {\em IEEE communications magazine}, vol.~55, no.~8, pp.~138--145, 2017.

\bibitem{popovski20185g}
P.~Popovski, K.~F. Trillingsgaard, O.~Simeone, and G.~Durisi, ``{5G wireless network slicing for eMBB, URLLC, and mMTC: A communication-theoretic view},'' {\em Ieee Access}, vol.~6, pp.~55765--55779, 2018.

\bibitem{yin2023connectivity}
B.~Yin, J.~Tang, and M.~Wen, ``Connectivity maximization in non-orthogonal network slicing enabled industrial internet-of-things with multiple services,'' {\em IEEE Transactions on Wireless Communications}, vol.~22, no.~8, pp.~5642--5656, 2023.

\bibitem{lien2017efficient}
S.-Y. Lien, S.-C. Hung, D.-J. Deng, and Y.~J. Wang, ``Efficient ultra-reliable and low latency communications and massive machine-type communications in {5G} new radio,'' in {\em GLOBECOM 2017-2017 IEEE Global Communications Conference}, pp.~1--7, IEEE, 2017.

\bibitem{blanquez2016eolla}
F.~Blanquez-Casado, G.~Gomez, M.~d.~C. Aguayo-Torres, and J.~T. Entrambasaguas, ``eolla: an enhanced outer loop link adaptation for cellular networks,'' {\em EURASIP Journal on Wireless Communications and Networking}, vol.~2016, pp.~1--16, 2016.

\bibitem{lai2015path}
I.-W. Lai, C.-H. Lee, K.-C. Chen, and E.~Biglieri, ``Path-permutation codes for end-to-end transmission in ad hoc cognitive radio networks,'' {\em IEEE Transactions on Wireless Communications}, vol.~14, no.~6, pp.~3309--3321, 2015.

\bibitem{park2020extreme}
J.~Park, S.~Samarakoon, H.~Shiri, M.~K. Abdel-Aziz, T.~Nishio, A.~Elgabli, and M.~Bennis, ``Extreme {URLLC}: Vision, challenges, and key enablers,'' {\em arXiv preprint arXiv:2001.09683}, 2020.

\bibitem{tang2024learn}
J.~Tang, F.~Chen, J.~Li, and Z.~Liu, ``Learn to schedule: Data freshness-oriented intelligent scheduling in industrial iot,'' {\em IEEE Transactions on Cognitive Communications and Networking}, 2024.

\bibitem{meredith2015study}
J.~M. Meredith, ``Study on downlink multiuser superposition transmission for {LTE},'' in {\em TSG RAN Meeting}, vol.~67, 2015.

\bibitem{you2016energy}
C.~You, K.~Huang, H.~Chae, and B.-H. Kim, ``Energy-efficient resource allocation for mobile-edge computation offloading,'' {\em IEEE Transactions on Wireless Communications}, vol.~16, no.~3, pp.~1397--1411, 2016.

\bibitem{destounis2018scheduling}
A.~Destounis, G.~S. Paschos, J.~Arnau, and M.~Kountouris, ``Scheduling {URLLC} users with reliable latency guarantees,'' in {\em 2018 16th International Symposium on Modeling and Optimization in Mobile, Ad Hoc, and Wireless Networks (WiOpt)}, pp.~1--8, IEEE, 2018.

\bibitem{destounis2019complexity}
A.~Destounis and G.~S. Paschos, ``Complexity of {URLLC} scheduling and efficient approximation schemes,'' in {\em 2019 International Symposium on Modeling and Optimization in Mobile, Ad Hoc, and Wireless Networks (WiOPT)}, pp.~1--8, IEEE, 2019.

\bibitem{lovasz2009matching}
L.~Lov{\'a}sz and M.~D. Plummer, {\em Matching Theory}, vol.~367.
\newblock American Mathematical Soc., 2009.

\bibitem{schulman2017proximal}
J.~Schulman, F.~Wolski, P.~Dhariwal, A.~Radford, and O.~Klimov, ``Proximal policy optimization algorithms,'' {\em arXiv preprint arXiv:1707.06347}, 2017.

\bibitem{yu2021surprising}
C.~Yu, A.~Velu, E.~Vinitsky, Y.~Wang, A.~M. Bayen, and Y.~Wu, ``The surprising effectiveness of {MAPPO} in cooperative, multi-agent games.,'' {\em arXiv preprint arXiv:2103.01955}, 2021.

\bibitem{haarnoja2018soft}
T.~Haarnoja, A.~Zhou, P.~Abbeel, and S.~Levine, ``Soft actor-critic: Off-policy maximum entropy deep reinforcement learning with a stochastic actor,'' in {\em International conference on machine learning}, pp.~1861--1870, PMLR, 2018.

\bibitem{schulman2015high}
J.~Schulman, P.~Moritz, S.~Levine, M.~Jordan, and P.~Abbeel, ``High-dimensional continuous control using generalized advantage estimation,'' {\em arXiv preprint arXiv:1506.02438}, 2015.

\bibitem{sinha2020d2rl}
S.~Sinha, H.~Bharadhwaj, A.~Srinivas, and A.~Garg, ``{D2rl}: Deep dense architectures in reinforcement learning,'' {\em arXiv preprint arXiv:2010.09163}, 2020.

\bibitem{qi2024efficient}
S.~Qi, L.~Lu, F.~Ziruo, B.~Xingzi, L.~Huaqiu, X.~Zhang, C.~Wen, and Y.~Jinpei, ``Efficient and fair ppo-based integrated scheduling method for multiple tasks of satech-01 satellite,'' {\em Chinese Journal of Aeronautics}, vol.~37, no.~2, pp.~417--430, 2024.

\bibitem{3GPP}
``Further advancements for {E-UTRA} physical layer aspects ({R}elease 9), {3GPP} standard {TS} 36.814,'' Mar. 2010.

\bibitem{kingma2014adam}
D.~P. Kingma and J.~Ba, ``Adam: A method for stochastic optimization,'' {\em arXiv preprint arXiv:1412.6980}, 2014.

\bibitem{arpaci2014chapter}
R.~H. Arpaci-Dusseau and A.~C. Arpaci-Dusseau, ``Chapter: Scheduling introduction,'' {\em Operating Systems: Three Easy Pieces; Arpaci-Dusseau Books: WI, USA}, 2014.

\end{thebibliography}

\end{document}